\documentclass{Interspeech2024}
\usepackage{xcolor}

\interspeechcameraready

\title{PLDNet: \textbf{P}LD-Guided \textbf{L}ightweight \textbf{D}eep \textbf{Net}work Boosted by Efficient Attention for Handheld Dual-Microphone Speech Enhancement}

\name[affiliation={1}]{Nan}{Zhou}
\name[affiliation={1}]{Youhai}{Jiang}
\name[affiliation={1}]{Jialin}{Tan}
\name[affiliation={1}]{Chongmin}{Qi}

\address{
  $^1$Shenzhen Transsion Holdings Co., Ltd, Shanghai Branch, China
}
\email{\{nan.zhou, youhai.jiang, jialin.tan, chongming.qi\}@transsion.com}


\begin{document}

\maketitle

\begin{abstract}
    
    Low-complexity speech enhancement on mobile phones is crucial in the era of 5G. Thus, focusing on handheld mobile phone communication scenario, based on power level difference (PLD) algorithm and lightweight U-Net, we propose PLD-guided lightweight deep network (PLDNet), an extremely lightweight dual-microphone speech enhancement method that integrates the guidance of signal processing algorithm and lightweight attention-augmented U-Net. For the guidance information, we employ PLD algorithm to pre-process dual-microphone spectrum, and feed the output into subsequent deep neural network, which utilizes a lightweight U-Net with our proposed gated convolution augmented frequency attention (GCAFA) module to extract desired clean speech. Experimental results demonstrate that our proposed method achieves competitive performance with recent top-performing models while reducing computational cost by over 90\%, highlighting the potential for low-complexity speech enhancement on mobile phones.
\end{abstract}

\section{Introduction}
\label{sec:intro}

Noise reduction represents a thoroughly explored domain within signal processing, characterized by a rich tapestry of prior research. Numerous conventional monaural methods have been developed over the years, utilizing robust noise estimation models~\cite{martin1994spectral,cohen2001speech,cohen2003noise}. However, dual-microphone speech enhancement methods demonstrate greater robustness in addressing transient noise and interference concurrently with stationary noise. Thus, they have attracted significant research attention. For example, researchers have proposed and developed power level difference (PLD) algorithms, which leverage the power variance of received dual-microphone signals to mitigate both stationary and non-stationary noise~\cite{yousefian2009using,jeub2012noise,zhang2012fast,fu2013dual}. Additionally, coherence function based approaches have proven popular for suppressing noise and interference~\cite{yousefian2011dual,yousefian2012dual}. Other widely used techniques in dual-microphone speech enhancement include beamforming-based methods such as minimum variance distortionless response (MVDR)~\cite{frost1972algorithm,higuchi2016robust} and generalized sidelobe canceler (GSC)~\cite{buckley1986adaptive}.

Over the past decade, there has been a proliferation of deep learning based monaural speech enhancement algorithms~\cite{xu2014regression,choi2019phase,hu2020dccrn,hao2021fullsubnet,zhang2022multi} which have achieved state-of-the-art performance. For dual-microphone speech enhancement, Tan et al. explored a real-time approach using densely-connected convolutional recurrent network (DC-CRN) and obtained notable results~\cite{tan2021deep}. Subsequently, Xu et al. further advanced the field by utilizing multi-head cross-attention (MHCA) mechanism~\cite{xu2022improving}. Additionally, we observe that the multi-scale temporal frequency convolutional network with axial attention (MTFAA-Net) demonstrated remarkable performance in the L3DAS22 3D speech enhancement challenge~\cite{zhang2022multi}.

Recently, there has been a trend to combine signal processing and deep learning approaches for speech enhancement. For example, an adaptive linear filter is often placed before a deep neural network (DNN) to perform acoustic echo cancellation~\cite{zhang2022multi}. Additionally, a hybrid approach integrating a Kalman filter and a recurrent neural network (RNN) was proposed by~\cite{yang2023low} and demonstrated greatly improved echo cancellation performance with low computational cost. Furthermore, an iterative refinement approach combining a DNN and a multi-frame multi-channel Wiener filter (mfMCWF) achieved first place in the L3DAS22 3D speech enhancement challenge~\cite{lu2022towards}. Compared to pure deep learning approaches~\cite{zhang2022multi,li2022embedding}, these hybrid techniques can promise better or competitive performance with reduced computational complexity, owing to the integration of signal processing priors.

In this study, for the handheld mobile phone communication scenario, we propose a low-complexity dual-microphone speech enhancement method by integrating signal processing and deep learning approaches. For this application scenario, the vocal component captured by the primary microphone is substantially stronger than that of the secondary microphone. For the signal processing guidance, we employ a PLD-based algorithm to pre-process the dual-microphone spectrum. We then utilize a lightweight U-Net with our proposed gated convolution augmented frequency attention (GCAFA) module to extract the desired clean speech. Experimental results demonstrate that our proposed method can achieve competitive performance with recent top-performing models while reducing computational cost by over 90\%. This highlights the potential of PLDNet for low-complexity speech enhancement on mobile phones.


\section{PLDNet}
\label{sec:method}

\label{ssec:pldnet}
\begin{figure*}[htbp]
\begin{center}
\includegraphics[width=0.9\linewidth]{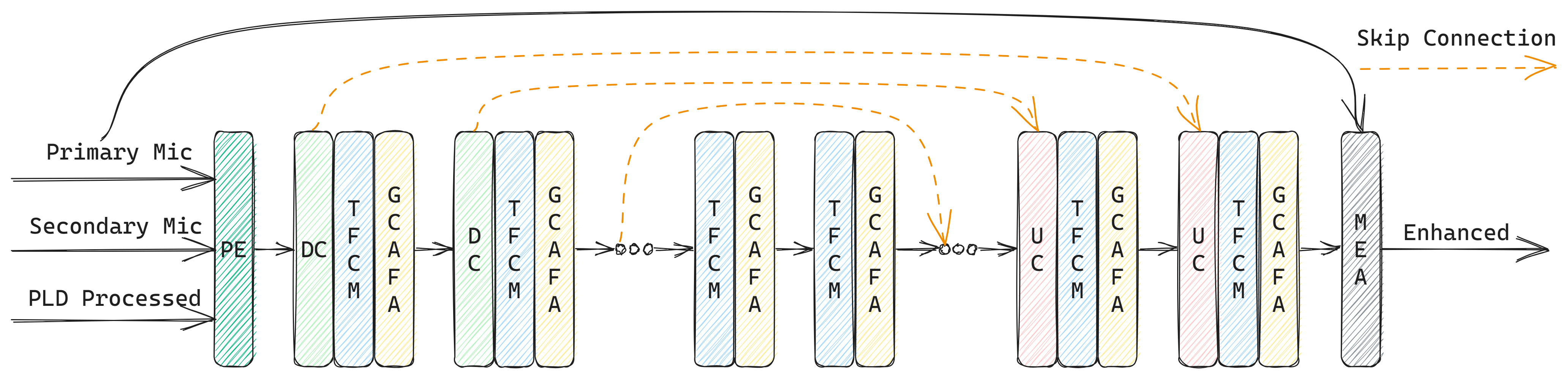}
\end{center}
\caption{Diagram of PLDNet.}
\label{fig:pldnet}
\end{figure*}

\subsection{Signal Model}
\label{ssec:signalmodel}

We define the signal model in the time-frequency (T-F) domain as:
\begin{equation}
\label{eq:signalmodel}
\resizebox{.8\hsize}{!}{$\begin{aligned}
Y_m(k,\ell) & = X_m(k,\ell) + V_m(k,\ell) + U_m(k,\ell) \\ & = X_m(k,\ell) + N_m(k,\ell), (m=1,2)
\end{aligned}$}
\end{equation}
where $Y_m(k,\ell)$, $X_m(k,\ell)$ and $N_m(k,\ell)$ are spectral coefficients of received, desired and undesired signals, respectively. $V_m(k,\ell)$ and $U_m(k,\ell)$ are stationary noise signal and transient interference signal. Furthermore, $k$ and $\ell$ are the frequency and time indices, whereas $m=1$ corresponds to the primary microphone and $m=2$ to the secondary microphone.

\subsection{PLD-based Pre-processing}
\label{ssec:preprocessing}

Drawing conceptual inspiration from GSC postfiltering algorithm~\cite{cohen2003analysis}, our PLD-based pre-processing algorithm consists of two main components: a stationary noise estimator~\cite{cohen2003noise} and a dual-channel collaborative noise reduction module. After obtaining an estimation for the stationary noise power spectrum $\lambda_m(k, \ell)$ using the noise estimation algorithm~\cite{cohen2003noise}, we calculate the posterior SNR $\gamma_m(k, \ell)$ of each microphone as:
\begin{equation}
\label{eq:PLD-1}
\gamma_m(k, \ell)=\frac{\left|Y_m(k, \ell)\right|^2}{\lambda_m(k, \ell)}
\end{equation}

Then we compute the ratio between the energy of the voice signals obtained from two microphones as:
\begin{equation}
\label{eq:PLD-2}
\kappa(k, \ell)=\frac{{\left|Y_1(k, \ell)\right|^2}-{\lambda_1(k, \ell)}}{{\left|Y_2(k, \ell)\right|^2}-{\lambda_2(k, \ell)}}
\end{equation}

The speech presence probability (SPP) is defined as:
\begin{equation}
\resizebox{.9\hsize}{!}{$
\begin{aligned}
\label{eq:PLD-3}
\psi(k, \ell)= 
\begin{cases}
1, & \text{if} \  \gamma_1(k, \ell)>\gamma_\text{thresh} \ \text{and}\\
& \ \kappa(k,\ell) > \kappa_\text{high} \\ 
\dfrac{\kappa(k, \ell)-\kappa_\text{low}}{\kappa_\text{high}-\kappa_\text {low}}, & \text{if} \ \gamma_1(k, \ell)>\gamma_\text{thresh} \ \text{and}\\
& \ \kappa_\text{low}<\kappa(k,\ell)<\kappa_\text{high}, \\ 
0, & \text {otherwise}\
\end{cases}
\end{aligned}$}
\end{equation}

\begin{table}
\begin{center}
\caption{Values of the constants used in the PLD-based pre-processing algorithm.}
\resizebox{0.9\linewidth}{!}{
\begin{tabular}{cccc}
\toprule
$k_{low} = 8$ & $k_{high} = 113$ & $\gamma_{thresh} = 1.69$ & $\tilde{\psi}_{thresh} = 0.25$ \\ $\kappa_{low} = 1.5$ & $\kappa_{high} = 3$ & $\gamma_{low} = 1$ & $\gamma_{high} =4.6$ \\
\bottomrule
\end{tabular}
}
\label{tab:PLD_contants}
\end{center}
\end{table}


Global SPP $\tilde{\psi}(\ell)$ is defined as:
\begin{equation}
\label{eq:PLD-4}
\tilde{\psi}(\ell)=\dfrac{1}{k_{high}-k_{low}+1} \sum_{k=k_{low}}^{k_{high}} \psi(k, \ell)
\end{equation}

Then $\gamma_1(k,\ell)$, $\psi(k, \ell)$ and $\tilde{\psi}(\ell)$ are used to compute the a posterior signal absence probability $\hat{q}(k, \ell)$ defined as:
\begin{equation}
\resizebox{.9\hsize}{!}{$
\label{eq:PLD-5}
\hat{q}(k,\ell)=
\left\{\begin{array}{l}
1, \ \ \ \ \text{if} \ \gamma_1(k,\ell) \leqslant \gamma_{\text {low }}\text{or} \ \ \tilde{\psi}(\ell) \leqslant \tilde{\psi}_{thresh}, \\
\max( \dfrac{\gamma_{\text {high }}-\gamma_1(k,\ell)}{\gamma_{\text {high }}-\gamma_{\text {low }}}, 1-\psi(k,\ell)), \text { otherwise }
\end{array}
\right.$}
\end{equation}
where $\tilde{\psi}_{thresh}$ is a predetermined threshold.

Then gain function $G(k,\ell)$ is obtained using the optimally modified log-spectral amplitude (OMLSA)~\cite{cohen2001speech} method. It is subsequently applied to the primary microphone spectrum as:
\begin{equation}
  \label{eq:PLD-6}
\hat{X}_{\mathrm{PLD}}(k, \ell) = G(k,\ell)Y_1(k,\ell)
\end{equation}
The resulting predicted target speech spectrum $\hat{X}_{\mathrm{PLD}}(k, \ell)$ is then utilized as a guidance feature for subsequent speech enhancement model. The constants employed in the PLD-based pre-processing algorithm are presented in Table~\ref{tab:PLD_contants}.

\subsection{GCAFA-Boosted Lightweight U-Net}
\subsubsection{Main structure}
\label{sssec:main_structure}

In this study, we leverage the speech enhancement capability of multi-scale multi-resolution U-Net architecture~\cite{hu2020dccrn,zhang2022multi,tan2021deep} along with our proposed GCAFA module. Specifically, we utilize the same main structure as MTFAA-Net~\cite{zhang2022multi}, which represents the state-of-the-art real-time monaural speech enhancement model.

As depicted in Figure~\ref{fig:pldnet}, the Phase Encoder (PE) initiates the process by extracting complex information from three inputs, producing real-valued feature maps. This is enhanced by the inclusion of spectrum from the PLD pre-processing technique, which assists in learning directional cues. The architecture further comprises layers for downsampling convolution (DC) and upsampling convolution (UC), Temporal Frequency Convolution Modules (TFCMs), the novel GCAFA modules, and a deep filter-based Mask Estimating and Applying (MEA) module, interconnected by skip connections for feature integration. Each encoder block starts with a DC layer, followed by a TFCM and a GCAFA module for comprehensive temporal and frequency analysis. Following the last encoder block, two backbone blocks incorporate TFCMs and GCAFA modules. Each Decoder blcok contains a UC layer, a TFCM, a GCAFA module, and a skip connection. The MEA module employs a 2-stage approach to refine speech masks, initially leveraging a deep filter to harness adjacent frequency and temporal information for magnitude mask estimation, subsequently enhancing phase mask precision to uplift speech quality. Further insights into the PE, DC, UC, TFCM, and MEA components are available in~\cite{zhang2022multi}, with our specific parameters outlined in Section~\ref{ssec:experimental_setup}. The GCAFA module is elaborated in the subsequent section.

\subsubsection{GCAFA}
\label{sssec:GCAFA}

\begin{figure}[htbp]
\centering
\includegraphics[width=0.85\linewidth]{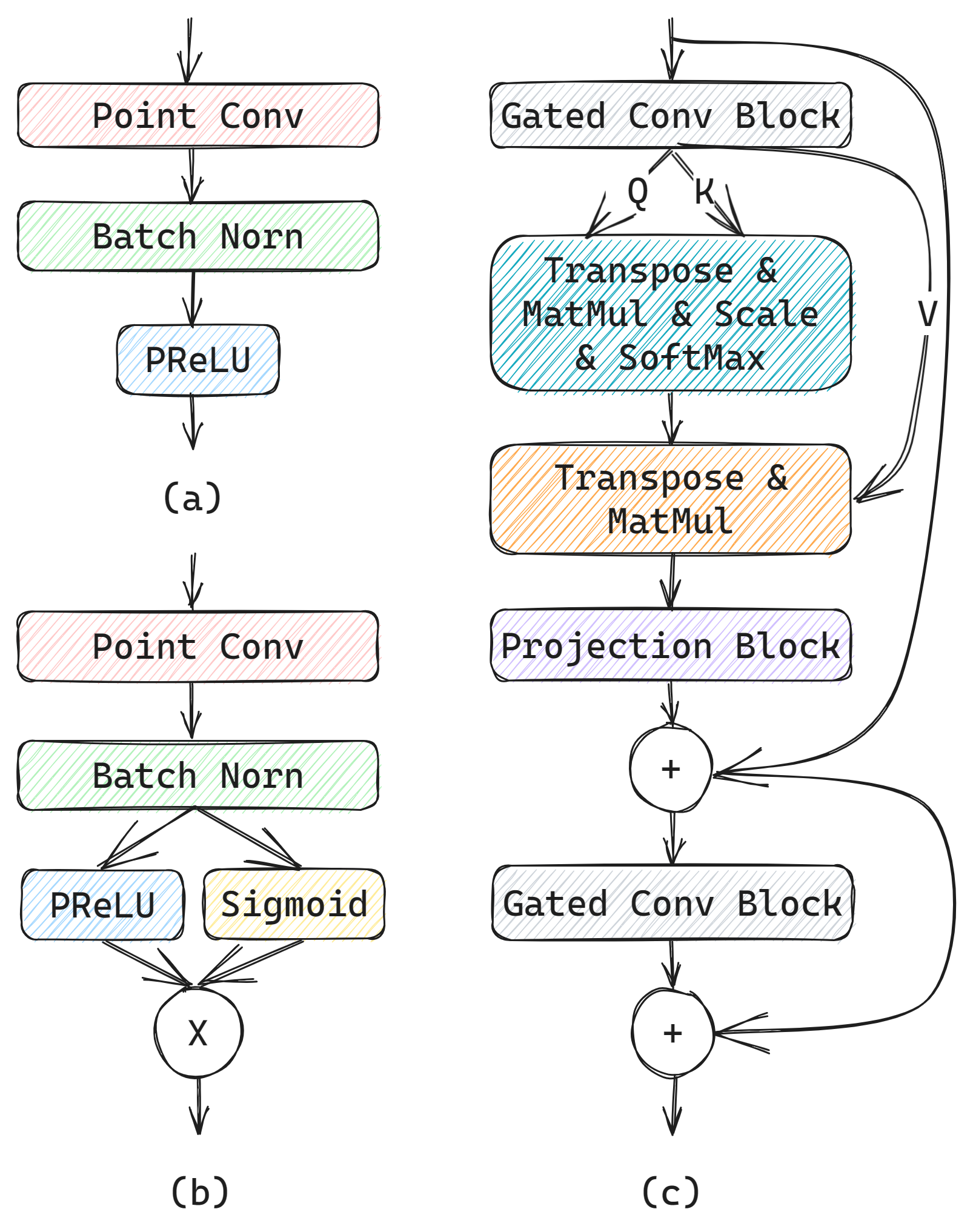}
\caption{Flow diagrams of projection block (a), gated convolution block (b), and GCAFA (c).}
\label{fig:GCAFA_diagram}
\end{figure}

GCAFA module integrates a low-complexity frequency attention mechanism, specifically designed to augment the model's capacity for speech processing. A pivotal observation underpinning our design is that while multi-scale dilated convolution blocks proficiently handle extended temporal contexts, they are inherently limited in addressing frequency-related nuances due to their constrained frequency kernel sizes.

To mitigate this limitation, the GCAFA module is engineered to adeptly capture and interpret global frequency dependencies. As illustrated in Figure~\ref{fig:GCAFA_diagram}, it incorporates a single-head self-attention mechanism that operates specifically along the frequency axis, ensuring a focused and effective analysis of frequency components. Additionally, pointwise convolution layers are utilized in both the input and output stages to encourage information exchange between channels. Finally, a gating mechanism is strategically implemented to regulate the flow of information.

The operational workflow of the GCAFA module begins with the input feature map, denoted as $ D \in \mathbb{R}^{B \times C \times F \times T} $. This feature map undergoes initial processing in a gated convolution block, which projects it onto a dimension of $ 3 \times C' $ channels. The resultant feature map is then partitioned into three segments along the channel dimension. At this juncture, the single-head self-attention mechanism comes into play, operating along the frequency axis to refine the frequency-specific features. Following the attention mechanism, the feature map's channel number is restored to $ C $ through a projection block. Finally, The process concludes with another gated convolution block. Notably, two residual connections are incorporated into the GCAFA module to facilitate training process and ensure stable convergence.

\subsubsection{Loss function}
\label{sssec:loss_function}

We use hybrid loss function combing waveform and spectral loss functions. The waveform loss function is defined as:
\begin{equation}
\label{eq:mtfcn-1}
L_{\mathrm{wave}} = \frac{\sum_{i=1}^{N} \left| \hat{x}_1(i) - x_1(i) \right|}{\sum_{i=1}^{N} \left| x_1(i) \right|}
\end{equation}
where $N$ is the number of samples in the waveform, $\hat{x}_1(i)$ denotes the predicted waveform and $x_1(i)$ denotes the ground truth waveform. The spectral loss function is defined as:
\begin{equation}
\label{eq:mtfcn-2}
\resizebox{.9\hsize}{!}{$
\begin{aligned}
L_{\mathrm{spec}} = \sum_{K \in 2^6, \ldots, 2^{11}} ( & \frac{\sum_{k} \sum_{\ell} ||\hat{X}_1(k,\ell) - X_1(k,\ell)||^2}{\sum_{k} \sum_{\ell}||X_1(k,\ell)||^2} + \\ & \sum_{k} \sum_{\ell} \frac{  ||\hat{X}_1(k,\ell) - X_1(k,\ell)||^2}{||X_1(k,\ell)||^2} )
\end{aligned}$}
\end{equation}
where $\hat{X}_1(k,\ell)$ and $X_1(k,\ell)$ denotes the predicted and ground truth spectra, $K$ is the frequency resolution.

The final loss function is defined as:
\begin{equation}
\label{eq:mtfcn-4}
L = L_{\mathrm{wave}} + \alpha L_{\mathrm{spec}}
\end{equation}
where $\alpha$ is the weight of $L_\mathrm{spec}$.

\section{Experiments}
\label{sec:experiments}

\subsection{Experimental Setup}
\label{ssec:experimental_setup}

In this study, we utilize three datasets for our experimental analysis, including VCTK~\cite{veaux2017cstr} for source speakers, LibriSpeech~\cite{panayotov2015librispeech} for interference speakers and WHAM!~\cite{wichern2019wham} for noise signals. From the VCTK dataset, we use 99, 5, and 5 speakers for training, validation, and testing, respectively. For the LibriSpeech dataset, we use the train-clean-100, dev-clean and test-clean subsets for training, validation, and testing, respectively. For the WHAM! dataset, we employ 58, 10, and 9 hours of data for training, validation, and testing, respectively.

To generate simulated signals, similar to~\cite{tan2021deep}, we simulate a rectangular room of dimension $10 \times 7 \times 3 \ \mathrm{m}^3$, with the target speech source located at the room's center. The primary microphone is placed on the same horizontal plane with the source, with a distance ranging from 2 to 5 cm. The secondary microphone is placed above the primary microphone, with a distance of 15 cm between the two microphones and a zenith angle ranging from $0^\circ$ to $15^\circ$. Furthermore, we generate all room impulse responses (RIRs) using image method, with RT60 values ranging from 0.2 to 0.5 s.

To simulate babble interference, similar to~\cite{tan2021deep}, we randomly select 72 speech clips form 72 different speakers in LibriSpeech dataset, and place them on a horizontal circle centered at the primary microphone, with a radius of 3 m. To simulate diffuse noise, we choose 2 noise clips from WHAM! dataset and mix them up using the method detailed in~\cite{habets2008generating}. Subsequently, we create all mixtures with SNR and SIR ranging from 0 to 20 dB and volume levels ranging from -40 to -10 dB with a sampling rate of 16 kHz.

\begin{table*}[htbp]
  \begin{center}
  \caption{Performance comparison of different speech enhancement models and ablation study of PLDNet on simulated test dataset. The best results of all models are highlighted in bold. The best results of PLDNet are highlighted in bold italics.}
  \resizebox{0.9\linewidth}{!}{
    \begin{tabular}{llllllllll}
      \toprule
      \textbf{Model} & \textbf{\#Params (M)} & \textbf{FLOPs/s (G)} & \textbf{SI-SDR} & \textbf{NB-PESQ} & \textbf{WB-PESQ} & \textbf{STOI} & \textbf{SIG} & \textbf{BAK} & \textbf{OVRL} \\
      \midrule
      Unprocessed & - & - & 7.325 & 2.364 & 1.333 & 0.880 & 2.735 & 2.082 & 1.962 \\
      \midrule
      PLD & - & 0.002 & 12.643 & 2.743 & 1.805 & 0.885 & 3.130 & 3.203 & 2.529 \\
      \midrule
      MTFAA-Net (mic 1)~\cite{zhang2022multi} & 2.149 & 4.542 & 16.384 & 3.080 & 2.305 & 0.941 & 3.149 & 4.038 & 2.898 \\
      MTFAA-Net (2 mics)~\cite{zhang2022multi} & 2.149 & 4.549 & 19.107 & 3.262 & 2.603 & 0.963 & 3.186 & 4.034 & 2.926 \\
      DC-CRN~\cite{tan2021deep} & 10.810 & 24.054 & 20.525 & 3.323 & 2.764 & 0.967 & 3.270 & \textbf{4.042} & 2.999 \\
      MHCA-CRN~\cite{xu2022improving} & 50.553 & 38.351 & \textbf{20.571} & \textbf{3.361} & \textbf{2.788} & \textbf{0.969} & \textbf{3.306} & 3.993 & \textbf{3.009} \\ 
      DC-CRN (scale-down) & 0.228 & 0.349 & 16.472 & 3.008 & 2.204 & 0.940 & 2.997 & 4.013 & 2.747 \\
      MHCA-CRN (scale-down) & 0.585 & 0.497 & 16.860 & 3.118 & 2.398 & 0.946 & 3.051 & 3.987 & 2.786 \\
      \midrule
      PLDNet & 0.155 & 0.312 & \textbf{\textit{19.219}} & \textbf{\textit{3.294}} & 2.653 & \textbf{\textit{0.962}} & \textbf{\textit{3.203}} & 4.004 & \textbf{\textit{2.925}} \\
      \ \ - w/o GCAFA & 0.115 & 0.199 & 17.949 & 3.274 & \textbf{\textit{2.658}} & 0.959 & 3.148 & \textbf{\textit{4.018}} & 2.885 \\
      \ \ \ \ - w/o PLD spectrum & 0.115 & 0.194 & 17.879 & 3.192 & 2.501 & 0.958 & 3.144 & 4.014 & 2.881 \\
      \ \ \ \ \ \ - w/o secondary microphone & 0.115 & 0.191 & 15.880 & 2.992 & 2.208 & 0.933 & 3.093 & 4.030 & 2.841 \\
      \bottomrule
    \end{tabular}
  }
  \label{tab:compare_simu}
  \end{center}
  \end{table*}


For the model settings, the PE's complex convolutional layer has 4 output channels, followed by 3 encoder blocks each with a DC layer (kernel: (7, 1), stride: (4, 1)), a TFCM with depth of 6, and a GCAFA module. In each GCAFA module, $C'$ is half of $C$. The output channels of 3 DC layers are 16, 24, and 40 across blocks. Two backbone blocks include dual 6-layer TFCMs and a GCAFA module each. The decoder inversely mirrors the encoder, with MEA's deep filter size set at (1, 3). Loss weight $\alpha$ is set at 1.0. STFT is conducted wiht a 32 ms window and 16 ms hop. Training involves 60 epochs, NovoGrad optimizer, cosine annealing scheduler, a 3e-3 initial learning rate, and batch size of 16. Notably, PLDNet's unoptimized version exceeds 3x real-time on an MT6789 ARM CPU, underscoring its real-time viability.

For comparison with recent SOTA dual-microphone speech enhancement models, we reproduce MTFAA-Net~\cite{zhang2022multi}, DC-CRN~\cite{tan2021deep} and MHCA-CRN~\cite{xu2022improving} on the same dataset. To compare with lightweight models, we reduce the model sizes of DC-CRN and MHCA-CRN to a fair level and train the models on the same dataset.

\subsection{Results and Analysis}
\label{ssec:results}

For performance evaluation, we adopt seven evaluation metrics, namely SI-SDR~\cite{le2019sdr}, NB-PESQ, WB-PESQ~\cite{rix2001perceptual}, STOI~\cite{taal2010short}, and SIG, BAK, OVRL using DNSMOS P.835~\cite{reddy2022dnsmos}. Additionally, we compute the number of parameters and FLOPs per second for all models using DeepSpeed toolkit~\cite{rasley2020deepspeed}. We compare the performance of various recent models against PLDNet, and present an ablation study of PLDNet. All models are causal versions for fair comparison.


The analysis summarized in Table~\ref{tab:compare_simu} shows that evaluated speech enhancement models outperform the PLD method in simulated environments, highlighting the superiority of modern deep learning over traditional approaches. A comparison between the monaural and binaural MTFAA-Net versions emphasizes the importance of multi-microphone inputs, especially in complex auditory settings involving multiple interfering speakers. An evaluation of MTFAA-Net, DC-CRN, and MHCA-CRN reveals a link between model complexity and performance improvements, up to a certain limit.

Moreover, PLDNet marginally exceeds MTFAA-Net's dual-microphone setup in performance, utilizing less than 10\% of its parameters and computational resources. It also significantly surpasses scaled-down versions of DC-CRN and MHCA-CRN, with its metrics closely approaching those of DC-CRN, despite DC-CRN having roughly 70 times more parameters and FLOPs. MHCA-CRN, the largest model compared, leads in performance but requires over 100 times the parameters and FLOPs compared to PLDNet. This highlights the efficacy of combining signal processing principles with a lightweight, attention-enhanced U-Net framework, challenging much larger models. An ablation study further reveals the GCAFA module's substantial improvement in metrics like SI-SDR, underlining its role in enhancing model performance. Moreover, PLD spectrum guidance boosts NB-PESQ and WB-PESQ scores, indicating superior speech quality. This suggests significant performance gains from signal processing-based guidance without heavy computational demands. A noticeable gap exists between dual-microphone and monaural models, as seen in the final row. And it's pertinent to highlight that it represents a version of monaural MTFAA-Net with 20 $\times$ fewer FLOPs and devoid of the attention mechanism, which performs marginally inferior to the original monaural MTFAA-Net. This underlines the challenges monaural models face in dual-channel enhancement, particularly due to the lack of spatial cues, leading to minimal differentiation in performance among them.

\begin{figure}
\centering
\includegraphics[width=0.95\linewidth]{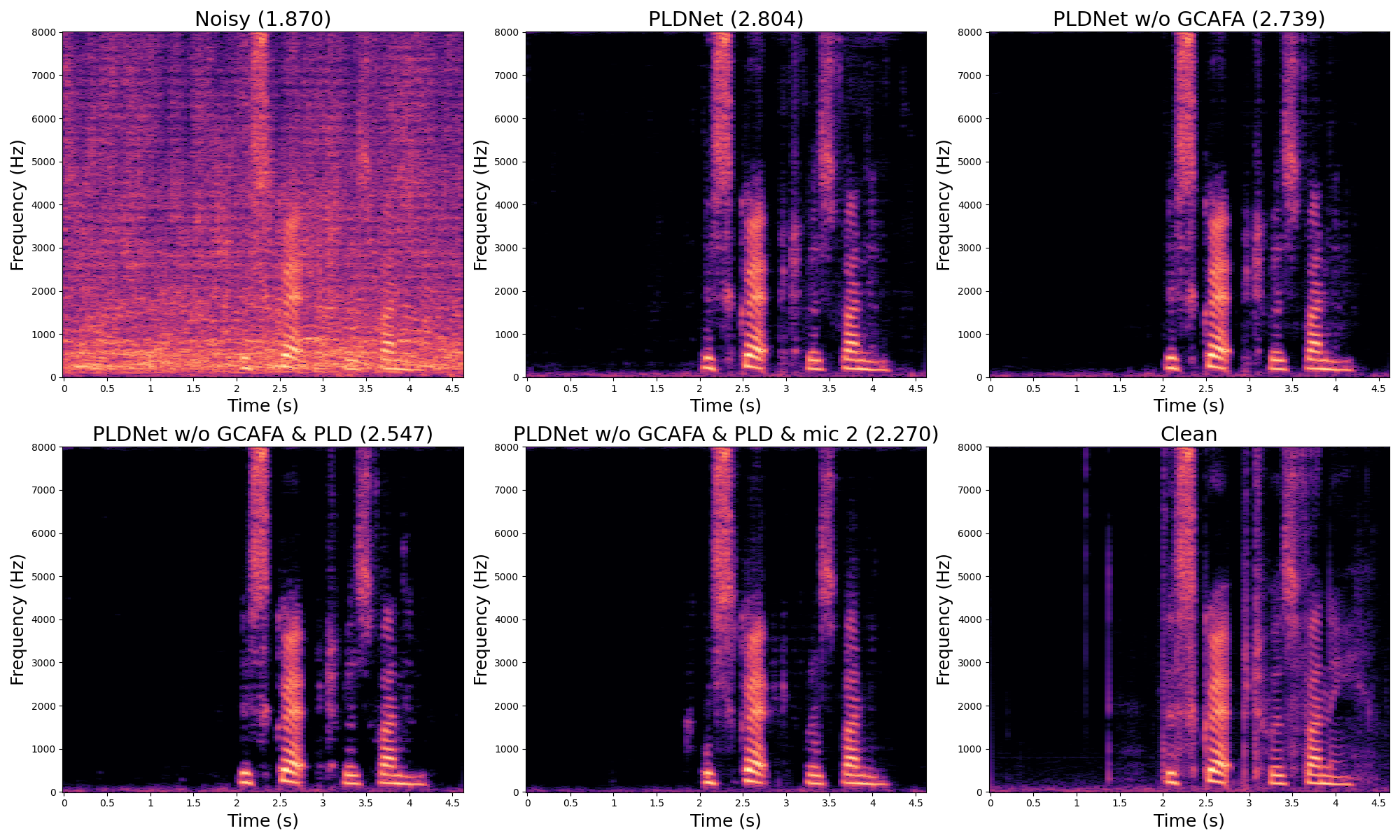}
\caption{Spectrograms from ablation study of PLDNet on a noisy sample, the corresponding NB-PESQ scores are displayed in parentheses.}
\label{fig:compare}
\end{figure}


Figure~\ref{fig:compare} presents spectrograms from an ablation study on PLDNet, applied to a noisy speech sample, alongside their corresponding NB-PESQ scores in parentheses. These results underscore PLDNet's effectiveness in minimizing background noise and maintaining essential speech elements. A noticeable difference in NB-PESQ scores between model versions with and without PLD spectrum guidance highlights its significance. The monaural model variant, however, exhibits considerable speech distortion, pointing out its deficiencies in safeguarding speech clarity.

\section{Conclusion}
\label{sec:conclusion}

In this study, we introduce PLDNet, an innovative, low-complexity speech enhancement approach that combines a PLD-based pre-processing technique with a streamlined U-Net architecture, enhanced by our GCAFA module. Targeting noise reduction in mobile phone communications, our method has shown to rival the performance of larger models with significantly lower computational demands. Future research will delve into merging signal processing with deep learning and the feasibility of real-time mobile deployment.

\bibliographystyle{IEEEtran}
\bibliography{mybib}

\end{document}